\documentclass[preprint2]{aastex}

\newcommand{\chg}[1]{#1}

\makeatletter

\newenvironment{inlinefigure}{%
\def\@captype{figure}%
\noindent\begin{minipage}{0.999\linewidth}\begin{center}}
{\end{center}\end{minipage}\smallskip}
\makeatother

\slugcomment{Accepted to ApJS}

\shorttitle{Angular Clustering in XMM-COSMOS}
\shortauthors{Miyaji et al.}

\begin{document}


%\title{Angular Clustering of the X-ray Sources in the First-year
%XMM-Newton Observations of the COSMOS Field
%\footnote{Based on observations using XMM-Newton Observatory.}}

\title{The {\sl XMM-Newton} wide-field survey in the COSMOS field: V. 
Angular Clustering of the X-ray Point Sources
\altaffiltext{$\star$}{Based on observations obtained with {\sl XMM-Newton}, 
an ESA science mission with instruments and contributions directly funded 
by ESA Member States and NASA.}}


\author{Takamitsu Miyaji\altaffilmark{1}, Giovanni Zamorani\altaffilmark{2},
 Nico Cappelluti\altaffilmark{3}, Roberto Gilli\altaffilmark{2},
 Richard E. Griffiths\altaffilmark{1}, Andrea Comastri\altaffilmark{2}, 
 G\"unther Hasinger\altaffilmark{3},  Marcella Brusa\altaffilmark{3}, 
 Fabrizio Fiore\altaffilmark{4},Simonetta Puccetti\altaffilmark{4},
 Luigi Guzzo\altaffilmark{5}, Alexis Finoguenov\altaffilmark{3}}

\altaffiltext{1}{Department of Physics, Carnegie Mellon University,
     5000 Forbes Avenue, Pittsburgh, PA 15213, USA}
\altaffiltext{2}{INAF-Osservatorio Astronomico di Bologna, via Ranzani 1, 
     I-40127 Bologna, Italy}
\altaffiltext{3}{Max-Planck Institut f\"ur extraterrestrische Physik,
     Postf. 1312, 85741 Garching, Germany}
\altaffiltext{4}{INAF-Osservatorio Astronomico di Roma, via Frascati 33, I-00100 Monteporzio, Italy}
\altaffiltext{5}{INAF-Osservatorio Astronomico di Brera, via Bianchi 46, I-23807 Merate (LC), Italy}
\begin{abstract}
We present the first results of the measurements of angular auto-correlation 
functions (ACFs) of X-ray point sources detected in the {\sl XMM-Newton} 
observations of the $\sim$ 2 deg$^2$ COSMOS field (XMM-COSMOS). \chg{A significant 
positive signals have been detected in the 0.5-2 (SFT) band,
in the angle range of 0.5-24 arcminutes, while the positive signals were at the  
$\sim$2 and $\sim 3\sigma$ levels in the 2-4.5 (MED) and 4.5-10 (UHD) keV bands
respectively.}  Correctly taking integral constraints into account is a major limitation in
interpreting our results.  With power-law fits to the ACFs
without the integral constraint term, we find correlation lengths of   
\chg{$\theta_{\rm c}=1\farcs 9\pm 0\farcs 3$, $0\farcs 8^{+0\farcs 5}_{-0\farcs 4}$ 
and $6\arcsec \pm 2\arcsec$} for the SFT, MED, and UHD bands respectively for $\gamma=1.8$. 
The inferred comoving correlation lengths, also taking into account the bias by the source 
merging due to {\sl XMM-Newton} PSF, are  \chg{$r_{\rm c}\approx 9.8\pm 0.7$, 
$5.8^{+1.4}_{-1.7}$ and $12\pm 2$ $h^{-1}$ Mpc at the effective redshifts of 
$\bar{z}_{\rm eff}\approx$ 1.1, 0.9, and 0.6} for the SFT, MED, and UHD bands 
respectively. If we include the integral constraint term in the fitting process,
assuming that the power-law extends to the scale length of the entire XMM-COSMOS
field, the correlation lengths become larger by \chg{$\sim 20$\%--90\%}. 
Comparing the inferred rms fluctuations of the spatial distribution of AGNs
$\sigma_{\rm 8,AGN}$ with those of the underlying mass, the bias parameters of the X-ray 
source clustering at these effective redshifts are in the range $b_{\rm AGN}=1.5-4$. 

\end{abstract}


\section{Introduction}

 Results from recent X-ray surveys have made very significant
contributions to understanding formation and evolution of supermassive
blackholes (SMBHs) at galaxy centers. In particular,
studies of X-ray luminosity function and its evolution have
been {providing the most reliable current estimates of} 
the accretion history to SMBH. One of the most important findings in 
recent years is that luminous active galactic nuclei (AGNs) arise earlier 
in the history of the universe than lower luminosity ones 
\citep{ueda03,SXLF3,barger05,lafranca05}. 
This suggests that more massive  SMBHs have been formed earlier in the 
universe, and reside quiescently at the centers of giant elliptical galaxies in the
later epochs, while more numerous, less massive SMBHs have been formed and 
accreted later in the history of the universe. 
 
%This is in the opposite sense to a semi-analytical model
%based on hierarchical structure formation picture, where accretion onto
%SMBH is related to merger events by \citet{wyithe}.
%However, N-body + Hydro cosmological simulations by \citet{dimatteo}
%\citep[see also][]{menci,} showed that accretion on black holes is 
%strongly linked to star formation and the presence of dense gas in the 
%center of galaxies. This resulted in a strong decline of luminous AGNs 
%towards lower redshift accompanied by the trend of low luminosity AGN 
%becoming more common later in the history of the universe, consistent 
%with the observation.

 Clustering properties of AGNs and their evolution with redshift provide yet 
additional clues to understanding {the accretion processes} onto the SMBHs. 
These give clues to environments of AGN activities.  
In the framework of the Cold Dark Matter (CDM) structure formation scenario, 
clustering properties or the bias of AGNs over a sufficiently large
scale  $b_{\rm AGN}=(\delta \rho/\rho)_{\rm AGN}/(\delta \rho/\rho)_{\rm mass}$
may be related to the typical mass of dark matter halos in which they reside,
\citep{mo96,sheth01}. At the same time, the mechanisms of triggering 
the AGN activity, which might be closely related to galaxy interactions and/or 
merging \citep{menci04,dimatteo05}, yield a clustering
of AGNs and can therefore be infered from the clustering analysis.   

 Since strong X-ray emission is a typical feature of an AGN activity, X-ray 
surveys provide most efficient means of constructing comprehensive 
complete samples of AGNs without contamination from the light in the stellar
population of host galaxies. In particular, surveys in the
harder ($E>2$ keV) X-ray band such as available from {\sl XMM-Newton}
are very efficient in finding not only unobscured AGNs, which are relatively
easy to select also in the optical bands, but also obscured ones, 
which are difficult to select with optical selection criteria alone. 
This is important because most of the accretion \citep[$\sim
80$\%;][]{coma95,gilli01,ueda03}
occurs in AGNs obscured by gas (in X-ray bands) and dust (in the 
optical bands). While one approach in investigating the environment of AGNs 
is to measure AGN overdensities  around known clusters of galaxies 
\citep{cappi01,delia04,cap05}, a more common and direct 
measure can be obtained by calculating auto-correlation functions (ACFs) of  
well-defined samples of X-ray selected AGNs.  

While small number statistics limits the accuracy of the 
clustering measurements of X-ray selected AGNs, there are a number 
of reports on the detection of the correlation signals. 
Samples based on the {\it ROSAT} 
All-Sky survey have mainly constrained the clustering properties of 
type 1 local AGNs at $z\la 0.3$. The correlation lengths resulting
from the angular \citep{akylas00} and 3D \citep{mullis04}
analyses of these samples are 6-7 $h^{-1}$ Mpc \footnote{Throughout this paper 
we adopt $(H_{\rm 0},\Omega_{\rm M},\Omega_\Lambda)=$
$(100\,h\;{\rm km\,s^{-1}\,Mpc^{-1}},0.3,0.7)$}. 
 Due to the wide redshift distribution,
it is more difficult to obtain clustering signal in deeper X-ray
surveys before redshifts for a complete set of X-ray sources 
are obtained. Nevertheless, \citet{basilakos04,basilakos05} measured 
strong angular correlation signals in their {\sl XMM-Newton}/2dF 
survey, which covers a total area of 2 deg$^2$ over two fields. They obtained 
a correlation length of $\sim 7.5\,h^{-1}$ Mpc in physical units for both soft
and hard X-ray selected sources, suggesting 
a clustering evolution which is fixed in the proper coordinate between
z$\sim 0$ and $z\gtrsim 1$. At much fainter X-ray fluxes, \citet{gilli05} 
analyzed the projected-distance correlation function $w(r_{\rm p})$
for the X-ray sources with spectroscopic redshifts in the Chandra Deep 
Fields North (CDF-N) and South (CDF-S). They found significantly 
different clustering properties in these two fields, suggesting a  
cosmic variance effect. Recently, \citet{yang06} made detailed 
analysis on their 0.4 deg$^2$ 
{\it Chandra} Large Area Synoptic X-ray Survey (CLASXS) supplemented
by CDF-N. With spectroscopic redshifts for a good portion of
the sources, they explored the clustering properties in different redshift and
luminosity bins as well as intrinsic absorption bins. They found the
evolution of bias with redshift but they did not find significant 
dependence in the clustering properties of X-ray selected AGNs based 
on either luminosity or intrinsic absorption.  
  
 One of the main aims of the COSMOS \citep{scoville06} project is 
to trace the evolution of the large-scale structure of the universe 
with an unprecedented accuracy and redshift baseline. The {\sl XMM-Newton}
Survey, covering the entire COSMOS field \citep[XMM-COSMOS;][]{hasinger06},
is one of the most extensive {\it XMM-Newton} Survey programs conducted 
so far. In the first-year {\sl XMM-Newton} observations, {about 1400} 
X-ray point sources 
have been detected and cataloged \citep{cap06} (hereafter C07), which are 
dominated by AGNs at redshifts $0.7<z<2$ \citep{brusa06,trump06}. 
 
In this paper, we report the first results of our investigations 
on the large scale structure  through an angular auto-correlation (ACF) 
function analysis of the X-ray point sources detected in XMM-COSMOS, as a 
preview of more detailed studies in the near future. Our future
studies include the derivation of the direct three-dimensional 
correlation function using redshift information already available for a large
portion of the X-ray sources and the analysis of the cross-correlation of 
the X-ray sources with galaxies in the multiwavelength COSMOS catalog.  

 The outline of the paper is as follows. In Sect. \ref{sec:sel}, we  
explain the selection of our samples of X-ray sources to be used in the 
correlation function analysis, which are subsets of those described in 
C07. Details of the calculations, including the ACF estimator,
the random sample, and power-law fits are presented in  
Sect. \ref{sec:acf}. The de-projection to the three-dimensional
correlation function is presented  in Sect. \ref{sec:3d}. The results
are discussed in Sect. \ref{sec:disc}. We summarize 
our conclusions in Sect \ref{sec:conc}

\section{Sample Selection}
\label{sec:sel}

Our samples consist of the X-ray sources detected in the
first-year {\sl XMM-Newton} observations of the COSMOS field. 
The source detection, construction of the sensitivity maps,
and source counts are described in C06.  The X-ray source catalogs in three
energy bands, corresponding to energy channels of 0.5-2 (SFT),
2-4.5 (MED) and 4.5-10 (UHD) keV are used. 

For the angular ACF studies, we have applied  further selection
criteria to the C06 sources to minimize the effects of possible systematic 
errors in the sensitivity maps. The applied criteria for this kind of 
analysis should be stricter 
than those adopted for the derivation of the $\log N - \log S$ function, 
because localized systematic errors may 
cause spurious clustering of X-ray sources. In order to do this, we have 
scaled up the original sensitivity map to:
\begin{equation}
CR_{\rm lim,acf} = \max (a\;CR_{\rm lim, C06} - b,\; CR_{\rm lim, min}),
\label{eq:limscal}
\end{equation} 
where $CR_{\rm lim, C06}$ is the limiting count rate ( in cts~s$^{-1}$) 
in the original C06 sensitivity map and $CR_{\rm lim,acf}$ is the sensitivity 
map used for the ACF analysis. After a number of trials, the scaling coefficients $(a,b)$ 
have been set to \chg{(1.33,$1\times 10^{-4}$), (1.40, 0.)
and (1.44, $4\times 10^{-4}$)} for the SFT, MED and UHD bands respectively. 
We have excluded the area where $CR_{\rm lim,acf}$
exceeds $CR_{\rm lim,max}$ (low exposure areas close to the field
borders).  Those X-ray sources with $CR$'s
below $CR_{\rm lim,acf}$ at the source position have been excluded from 
the ACF analysis. After these screenings, the numbers of sources for the ACF
analysis are \chg{1037, 545, and 151} for the SFT, MED, and UHD bands respectively. 
While the sensitivity in the soft band is the best among the
three bands, some X-ray sources are hard enough that they
are detected in only MED and/or UHD bands. These are mainly highly obscured 
AGNs. Out of the \chg{545 (151) MED (UHD) band sources after the screening
process, 59 (13)} have not been detected in the SFT band, and only one UHD sources 
have escaped the detection in the MED band \chg{(before the screening process)}.   
The numbers of the X-ray sources, values of $CR_{\rm lim,min}$,$CR_{\rm lim,max}$ 
and the total areas used for the ACF analysis are summarized in Table~\ref{tab:src}. 

\section{Angular Correlation Function Calculation}
\label{sec:acf}

\subsection{The ACF calculation}

 In calculating the binned ACF, we have used the standard estimator
by \citet{landy93}:
\begin{equation}
w_{\rm est}(\theta_i)=(DD-2DR+RR)/RR,
\label{eq:ls}
\end{equation}
where DD, DR, and RR are the normalized numbers of pairs in the $i$-th 
angular bin for the data-data, data-random, and random-random samples 
respectively. Also we use the symbols $D$ and $R$ to represent the 
data and random samples respectively. 
Expressing the actual numbers of pairs in these three
combinations as $n_{\rm pair, DD}(\theta_i)$, $n_{\rm pair, DR}(\theta_i)$ and 
$n_{\rm pair, RR}(\theta_i)$, the normalized pairs are expressed by:  
\begin{eqnarray}
  DD = n_{\rm pair,DD}(\theta_i)/[N_{\rm D}(N_{\rm D}-1)]\nonumber \\
  DR = n_{\rm pair,DR}(\theta_i)/(N_{\rm D}N_{\rm R})\nonumber\\
  RR = n_{\rm pair,RR}(\theta_i)/[N_{\rm R}(N_{\rm R}-1)] 
\end{eqnarray}
where $N_{\rm D}$, and $N_{\rm R}$ are the numbers of sources in the data and 
random samples respectively. 
The number of objects in the random sample has been set to 20 times of 
that in the data sample. This makes the variance of the second and third
terms of Eq. \ref{eq:ls} negligible in the error budget of $w_{\rm est}$.

%Thus a simple straight calculation of the ACF using the 
%detected sources should make a spurious correlation function reflecting
%the apparent clustering of sources due to the instrumental effects. One 
%approach is to make a uniform sample select sources with count rates ($CRs$) 
%above a certain cut $CR_{\rm lim}$. Also we only use the region where the 
%observation is sensitive enough to detect a source  with $CR_{\rm lim}$.
%However, with this procedure, one  should throw out a significant portion
%of the data and the loss of the correlation signal due to the reduced 
%sample can be critical for  relatively small numbers of objects
%detected in the X-ray surveys. 

 Our {\sl XMM-Newton} observations have varying sensitivity over the
field. In order to create a random sample, which takes the inhomogeneity 
of the sensitivity over the field into account, we have taken the 
following steps.
\begin{enumerate}
\item Make a random sample composed of $N_{\rm R}$ objects, where 
   $N_{\rm R}$ is an integer times $N_{\rm D}$. 
\item For each random object, assign a count rate from a source
  from the data sample. The assignments are made in sequence so that the CR
  distributions of the random and the data sample objects are exactly the same.
\item For each random object, assign a random position in the field. If the 
  sensitivity-map value at this position ($CR_{\rm lim,acf}$ from Eq.
\ref{eq:limscal}, 
  in units of counts s$^{-1}$) is larger 
  than the assigned CR, find a different position. Repeat this until the 
  position is sensitive enough to detect a point-source with the assigned 
  CR.
\end{enumerate}

 As a check on this procedure, we have also calculated ACFs using two other  
methods of generating random samples. The second method is to 
assign the count rates to the random sources drawn from a 
$\log N$ -- $\log S$ relation \citep[e.g.][]{moretti03},
instead of copying the count rates of the data sample. Then the source
is placed at a random position in the field. If the sensitivity limit
at this position is higher than the assigned CR, this random
source is rejected. Another, but more sophisticated and computationally 
demanding method is to generate random sources based on the $\log N$ -- $\log
S$ 
relation, down to a flux level much lower than the sensitivity limit of our 
observations. These sources are then fed into a simulator, taking into
account {\sl XMM-Newton} instrumental effects, including position-dependent 
PSF, exposure maps, and particle background. The entire first-year 
XMM-COSMOS image has been simulated and the same
source detection procedure as that applied to the actual data has been
applied on the simulated data. A random sample $R$ is generated from
10 simulated XMM-COSMOS fields. Using the above two methods for random sample 
generations did not alter the results significantly. In the following analysis, 
we show the results obtained by the first method. 

\subsection{Error Estimation and Covariance Matrix}

 We have estimated the errors using the variance of the calculated 
ACF by replacing $D$ in Eq. \ref{eq:ls} by random samples. A random 
sample with the same number of objects and the same set of count rates
as $D$ has been drawn independently from $R$ in Eq. \ref{eq:ls}. We denote the 
random sample as a replacement of $D$ during the error search by $R^\prime$ to 
distinguish from $R$. For each angular bin, a 1 $\sigma$ 
error has been calculated as a standard deviation from resulting
ACFs calculated from $N_{\rm run}=$80 different $R^\prime$ samples, which is 
then multiplied by a scaling factor $[1+w(\theta_i)]^{1/2}$ (hereafter referred
to
as ``the scaled random errors''). This scaling factor corrects for the
difference between
the errors in the null-hypothesis case, obtained from $R^\prime$s, and those
in the presence of the correlation signal. This is in line with the observation that
the error in each bin of the ACF calculated using the \citet{landy93} 
estimator is approximately a Poisson fluctuation of the number of 
data-data pairs in the bin, i.e., $\sigma \sim [1+w(\theta)]/\sqrt{n_{\rm pair,
DD}}$.  
The standard deviation of the null-hypothesis ACF, obtained by replacing $D$ by 
an $R^\prime$, is $\sigma_{\rm ran} \sim 1/\sqrt{n_{{\rm pair},R^\prime R^\prime}}$.
The 
scaling factor can be obtained by using the relation,
$n_{\rm pair,DD}\approx [1+w(\theta)]n_{\rm pair,R^\prime R^\prime}$.  

  In correlation functions, the errors in different angular bins are
not independent from one another and correlations among the errors have to 
be taken into account when we make function fits.  Thus we have also 
estimated full covariance matrix in order to represent the correlations among 
errors by
\begin{eqnarray}
M_{{\rm cov},ij}&=\sum _{k}(w_{{\rm R},k}(\theta_i)-\langle w_{\rm
R}(\theta_i)\rangle)\nonumber\\
 &\times (w_{{\rm R},k}(\theta_j)-\langle w_{\rm R}(\theta_j)\rangle)/N_{\rm
run}\nonumber\\
 &\times (1+w_{\rm est}(\theta_i))^{1/2}(1+w_{\rm est}(\theta_j))^{1/2}
\label{eq:cov}
\end{eqnarray}
where $w_{{\rm R},k}(\theta_i)$ is the ACF value for the $k$-th random run
($k$ runs through 1 to $N_{\rm run}=80$),  $\langle w_{\rm R}(\theta_i)\rangle$
is their mean value, evaluated at the center of the $i$-th angular bin 
$\theta_i$ and $w_{\rm est}(\theta_i)$ is from Eq. \ref{eq:ls}. The square
roots of the diagonal elements of Eq. \ref{eq:cov} are the scaled random errors
discussed above. The covariance matrix calculated in Eq. \ref{eq:cov} is used 
later in Sect. \ref{sec:fits}. Strictly speaking, Eq. \ref{eq:cov} only takes
into 
account the correlations of errors for the random cases, but not the 
correlation of errors due to clustering. One way of explaining this is that  
removing/adding one source (by a Poisson chance) affects multiple 
angular bins and this is represented by  non-diagonal
elements of Eq. \ref{eq:cov}. 
 On the other hand, correlation of errors due to large scale structure 
or the cosmic variance is not represented by this. If we observe another part 
of the sky, we sample different sets of large scale structures such as
filaments 
and voids. Since the existence or non-existence of one such structure affects 
multiple angular bins, there should be additional contribution to the
non-diagonal 
elements of the  correlations among errors in different angular bins. 
Eq. \ref{eq:cov}, based on many random samples, thus includes the former 
type of the correlation of errors but not the latter. One way to include 
also the latter effect is to use the Jackknife re-sampling technique,
as was done by e.g. \citet{zehavi04}. However, the Jackknife re-sampling
requires to divide the sample into many statistically independent
regions, which is not practically possible in our case.

\subsection{The binned ACF results}

\begin{figure*}
%\centerline{\includegraphics[width=\textwidth]{acf3_col.eps}}
\centerline{\includegraphics[width=\textwidth]{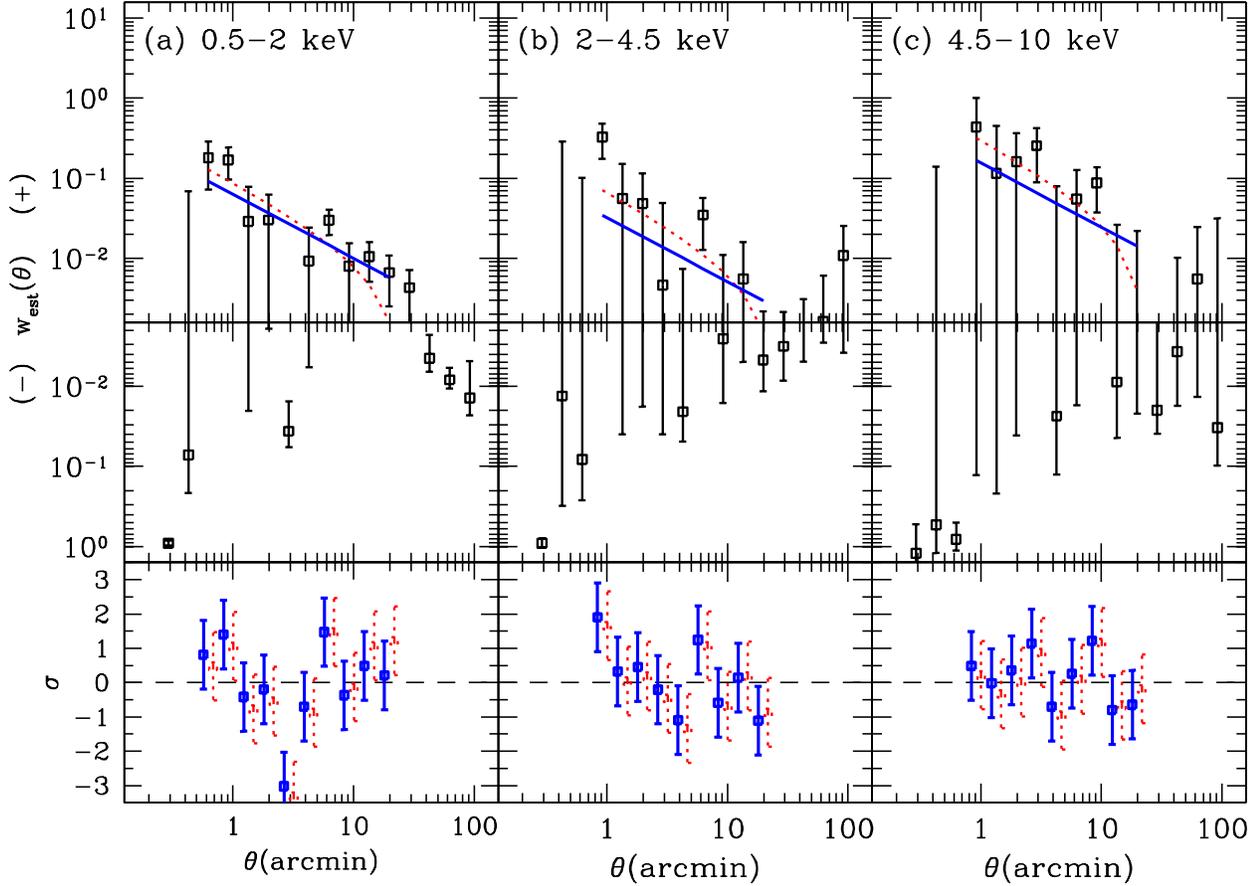}}
\caption
 { The binned estimated angular correlation functions $w_{\rm est}(\theta)$
are plotted for the X-ray sources detected in the first-year
{\sl XMM-Newton} data in three standard energy bands as labeled. The 
vertical scale is logarithmic, where the positive and negative 
parts ($\log |w_{\rm est}(\theta)|>-2.8$ respectively) are attached 
together. The 1$\sigma$ errors are the diagonal components of Eq. \ref{eq:cov}, 
i.e., the scaled random errors. The blue-solid and red-dotted lines (colors
in the electronic version only) show the best-fit power-law models 
for $\gamma-1=0.8$ without and with an integral constraint respectively. The
models
are plotted in the range where the fits are made. 
Fit residuals in terms of $\sigma$ has been also plotted in the lower 
panels for the two models (slightly offset for visibility) in the same
line styles (colors) as the models.
\label{fig:acf3}}
\end{figure*}

 The ACFs have been calculated for the three bands in logarithmically 
equally spaced bins with 
$\Delta \log\theta=1/6$. The results are shown in Fig \ref{fig:acf3},
where the upper panels, composed of two layers of logarithmic plots 
with positive and negative parts ($\log |w(\theta)|>-2.8$ 
respectively), are attached together. The lower panels show fit residuals
for the best-fit functions described in Sect. \ref{sec:fits}. Changing
the bin size did not change the clustering amplitude significantly.

The ACFs are presented with the scaled random errors. 
 Positive signals have been detected down to $\theta \sim 0.5\arcmin$ in
the 0.5-2 keV band and $\theta \sim 1\arcmin$ in the other bands. 
At the smallest scales, correlation signal is negative,
probably due to confusion effects, where two sources separated by a
distance comparable to or closer than the point spread function (PSF)
cannot be detected separately in the source detection procedure and
may well be classified as one extended source. In our current sample,
the sources that have been classified as extended have been
removed from the sample. The effect of this is discussed in detail 
in Sect. \ref{sec:merge}. 
In the 0.5-2 and 2-4.5 keV bands, positive
signals extend out to $\sim 20\arcmin$. 
Negative signals are seen at the largest angular scales, 
probably due to the integral constraint as discussed below.

%
% About the 1.5 arcmin dip...
%
%  Fig. \ref{fig:acf3} shows that there seems to be a dip in the ACF at 
%$1.5\arcmin \lesssim \theta \lesssim 3\arcmin$ in any band, although
%the significance of the dip is in any case smaller than 2$\sigma$. The 
%angle range is the place where the positive signal to noise ratio becomes
%poor by the interplay between the absolute value of ACF and the number
%of pairs. Thus the dip, which gives an enhanced impression in the 
%logarithmic plots, may well be caused by just a statistical fluctuation.   

\subsection{Power-law Fits}
\label{sec:fits}

  In order to make a simple characterization of our ACF results, we have 
fitted the ACF with a power-law model of the form:
\begin{equation}
w_{\rm mdl}(\theta)=A\;\theta^{1-\gamma}\nonumber
\label{eq:wthpl}
\end{equation}
where $\gamma$ is the slope index of the corresponding 
three-dimensional correlation function.
We use the normalization $A$ as a fitting parameter rather  
than the angular correlation length $\theta_{\rm c}=A^{1/(\gamma-1)}$,
since this gives much better convergence of the fit.
   
 The fits are made by minimizing $\chi_{\rm c}^2$. The subscript ${\rm c}$ 
denotes that the correlations between errors have been taken into account through 
the inverse of the covariance  matrix:
\begin{equation}
\chi_{\rm c}^2 ={\bf \Delta}^{\rm T}\,{\bf M}_{\rm cov}^{-1}{\bf\Delta},
\end{equation}
where ${\rm \Delta}$ is a vector composed of $w_{\rm est}(\theta_i)-w_{\rm mdl}(\theta_i)+C$, 
$M_{\rm cov}$ is the covariance matrix calculated in Eq. \ref{eq:cov} with $N_{\rm run}=80$,
and $C$ is a constant to compensate for the integral constraint as discussed below.

 Due to the finite area and the construction of $w_{\rm est}$ in
Eq. \ref{eq:ls}, the estimated angular correlation function
satisfies the integral constraint \citep[e.g][]{basilakos05,roche99}:
\begin{equation}
\int \int w_{\rm est} d^2\Omega = 0.
\label{eq:integ}
\end{equation}
 This constraint usually results in  $w_{\rm est}$ underestimating 
the true underlying angular correlation function by the constant $C$. 
Under an assumption that the true underlying 
$w(\theta)$ is a power-law and is extended to the scale of  
the survey area, one can include $C$ in the fitting process, where $C$ can be
uniquely determined by $\theta_{\rm c}$ and $\gamma$ by imposing the integral
constraint \citep{roche99},
\begin{equation}
C = \sum_i A\theta_i^{1-\gamma}\,RR(\theta_i)/\sum_i RR(\theta_i),
\end{equation} 
where the sums are over angular bins and $RR(\theta_i)$ is the number
of random-random pairs in the $i$-th angular bin. {The above assumption
is not necessarily true. If residual systematic errors in the sensitivity
maps are the main cause of the negative values at large angular separations,
the determination of $C$ shown above is not valid. However, this gives an 
approximate estimate of the degree of the underestimation by the integral
constraint.} 
This sets an limitation to the our angular ACF analysis, where 
the estimated $C$ values are not negligible compared with the amplitude 
of the ACF signal.  
We have made fits with or without including the integral constraint.

 Because of the limited signal-to-noise ratio, we were not able to
constrain $A$ and $\gamma$ simultaneously. Thus we have calculated
the best-fit values and 1$\sigma$ confidence errors for the amplitude 
for two fixed values of $\gamma-1=0.8$ and $0.5$. The former value
is for the canonical value for local galaxies
\citep[e.g.][]{peebles80,zehavi04},
the latter is approximately the slope found for X-ray sources in the 
{\sl Chandra} Deep Fields \citep{gilli05}. 
The angular fit results are summarized in Table \ref{tab:fit}. In this table, 
fits with different bands and parameters are identified with a Fit ID. 
\chg{The angle range for the fits are  $\theta_{\rm min}<\theta<\theta_{\rm max}$
and the boundaries are also shown in Table \ref{tab:fit}.   
For fit ID's S1-S4, M1-M4, and U1-U4, we have set $\theta_{\rm min}=0.5\arcmin$ (SFT) or 
$0.7\arcmin$ (MED,UHD), which is the minimum at which ACF is still positive, and below
which the ACF goes negative due to the {\sl XMM-Newton} PSF. Likewise, we set 
$\theta_{\rm max}=24\arcmin$, which is about the maximum scale where 
the ACF is still positive. The best-fit models for $\gamma-1=0.8$ 
are overplotted in Fig. \ref{fig:acf3} in the bin ranges included 
in the fits.} 

\chg{As another choice, we have set $\theta_{\rm min}$ and $\theta_{\rm max}$ 
in such a way that the range approximately corresponds to the projected comoving 
distance range of 1-16 $h^{-1}$ Mpc (Fit ID S5,S6,M5,M6,U5, and U6) at the
effective median redshift of the sample ($\bar{z}_{\rm eff}$ discussed below 
in Sect. \ref{sec:3d}.  The rationale for the maximum scale is that, 
in our subsequent analysis, the correlation functions are converted to the 
root mean square (rms) density fluctuation with a $8 h^{-1}$ Mpc-radius sphere 
(therefore the relevant maximum separation is $16 h^{-1}$) in discussing bias parameters. 
The rationale for the minimum scale is to minimize the effects of non-linearity 
in discussing the bias parameters and typical halo masses.}

\subsection{Effects of Source Merging due to PSF}
\label{sec:merge}

 The {\em amplification bias}, due to which the estimated ACF 
from sources detected in a smoothed image (e.g. by a finite PSF) is 
amplified with respect to the true underlying ACF, has been first
noted and discussed by \citet{vikhlinin95} in the context of the 
clustering of X-ray sources. This is caused by merging of multiple 
sources which are separated by distances comparable to or closer than 
the PSF. The effect of this bias depends on
the true underlying angular correlation function and the number density
of the sources. In principle, full simulations involving PSF smoothing and 
the source detection process are required to estimate the amount of this 
bias. \citet{basilakos05} took a simplified approach in estimating this 
effect on their ACF from their {\sl XMM-Newton}/2dF survey. In order 
to estimate the size of the effect, they used particles sampled from a
cosmological simulation. They simulated the {\sl XMM-Newton} sources by merging all 
the particle pairs closer than $6\arcsec$. They then compared the angular ACFs
from the particles themselves and the simulated {\sl XMM-Newton} sources. As a result,
they estimated that the measured angular correlation length is overestimated 
by  3-4\% due to the amplification bias.

 In our case with XMM-COSMOS, we have explicitly excluded sources that are
classified as extended by the source detection procedure (C06). This 
causes most of the source pairs closer than $\sim 20\arcsec$ to disappear from 
the sample, since these pairs are classified as single extended sources.
{Because the exclusion of these sources can 
suppress the estimated angular correlation function, we use the term 
``PSF merging bias'' rather than the ``amplification bias''.} 
Pairs of sources that are closer than $\sim 4$
arcseconds are, however, detected as a single point source. 
We have applied a similar approach to \citet{basilakos05} in estimating the 
effects {of the PSF merging bias}.   We have sampled 
particles from the COSMOS-Mock catalog extracted from the Millennium simulation
(Kitbichler, M., priv. comm.) over $\sim 2$ deg$^2$ of the sky. {Redshift, 
cosmological intrinsic redshift, and magnitudes in various photometric bands 
are provided for each mock galaxy in the catalog.} 
{We use the mock catalog to estimate the effect of
the PSF merging bias in the {\em angular correlation function}. 
Thus the selected objects from the mock catalog for our simulation do not have 
to physically represent to the actual X-ray selected AGNs.  
For our present purpose, we have chosen the mock galaxies in a redshift 
interval (roughly in the range $0.4\la z \la 0.8$) and  a magnitude range 
in such a way that the amplitude of the resulting angular ACF and the 
source number densities roughly match those of the X-ray samples.} 
We have then created a simulated 
XMM-COSMOS catalog as follows: 1) source pairs with separations smaller than $4$ 
arcseconds are merged into single sources and 2) pairs that are between 4 to 
20 arcseconds from each other are eliminated. We repeated this experiment 19 times 
and compared the mean angular 
ACFs from the original particles and that from simulated XMM-COSMOS by making
power-law fits to the mean ACFs. As a result we found that the ACF 
amplitudes measured using the sources in our source detection 
procedure on the XMM-COSMOS data are underestimated by 15\% and 8\% for the 
SFT and MED bands respectively. Corrections for this effect have not been
applied for the values in Table \ref{tab:fit}, but are considered in further 
discussions. The effects is negligible in the UHD band, due to the 
relatively low number density of the sources detected in this band, 
which made the average distance among neighboring sources much larger than the PSF. 

\section{Implication for 3-D Correlation Function and Bias}
\label{sec:3d}

\subsection{De-Projection to Real Space Correlation Function}

 The 2-D ACF is a projection of the real-space 3-D ACF 
of the sources $\xi(r)$ along the line of sight. In the following
discussions and thereafter, $r$ is in comoving coordinates. The
relation between the 2-D (angular) ACF and the 3-D ACF is expressed by 
the Limber's equation \citep[e.g.,]{peebles80}. Under the usual assumption
that the scale length of the clustering is much smaller
than the distance to the object, this reduces to:

\begin{eqnarray}
w(\theta)N^2= \int\left(\frac{dN}{dz}\right)^2 \times \nonumber \\ 
   \int \xi(\sqrt{[d_{\rm A}(z)\theta]^2+l^2}\;(1+z)) &
\left(\frac{dl}{dz}\right)^{-1}dl\;dz, 
\end{eqnarray}  
where $d_{\rm A}(z)$ is the angular distance, $N$ is the total
number of sources and $dN/dz$ is the redshift distribution (per $z$)
of the sources. The redshift evolution of the 3-D correlation
function is customarily expressed by 
\begin{equation}
\xi(r,z)=(r/r_{\rm c,0})^{-\gamma}(1+z)^{-3-\epsilon+\gamma},
\end{equation} 
where $\epsilon=-3$ and $\epsilon=\gamma-3$ correspond to the case 
where the correlation length is constant in physical and comoving
coordinates respectively. In these notations, 
the zero-redshift 3-D correlation length $r_{\rm c,0}$ can be related to 
the angular correlation length $\theta_{\rm c}$ by:
\begin{eqnarray}
&r_{\rm c,0}^\gamma = (N^2/S)\,\theta_{\rm c}^{\gamma-1},\nonumber \\
&S=H_\gamma \int \left(\frac{dN}{dz}\right)^2
  \left(\frac{c\,d\tau (z)}{dz}\right)^{-1}d_{\rm A}^{1-\gamma}
   (1+z)^{-3-\epsilon} dz \nonumber \\
&H_\gamma=\frac{\Gamma[(\gamma-1)/2]\Gamma(1/2)}{\Gamma(\gamma/2)},
\label{eq:proj}
\end{eqnarray}
where $\tau (z)$ is the look back time. Note that all dependence
on cosmological parameters are included in $d_{\rm A}(z)$ and
$\tau (z)$. We also define the comoving correlation length:
\begin{equation}
r_{\rm c}(\bar{z}_{\rm eff}) = r_{{\rm c},0}(1+\bar{z}_{\rm
eff})^{(-3-\epsilon+\gamma)/\gamma}
\end{equation}
at the effective redshift $\bar{z}_{\rm eff}$, which is the
median redshift of the contribution to the angular correlation (the integrand
of the second of Eq. \ref{eq:proj}).

\begin{inlinefigure}
%\centerline{\includegraphics[width=\textwidth]{zdis.eps}}
\centerline{\includegraphics[width=\textwidth]{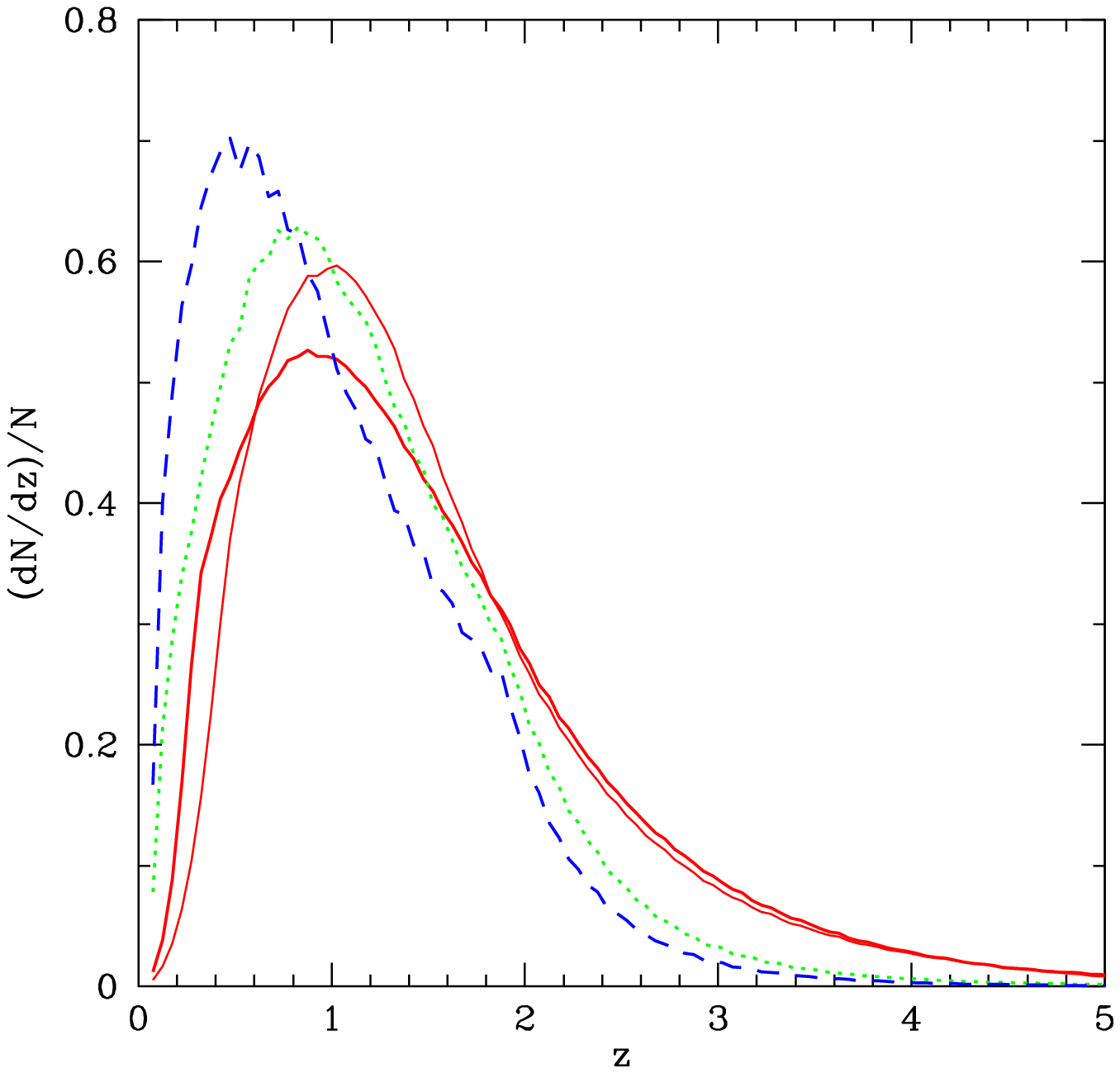}}
\caption
 {The model redshift distributions of the COSMOS-XMM sources 
in the 0.5-2 keV (solid lines), 2-4.5 keV (dotted line), 
and 4.5-10 keV (dashed line) respectively. The thicker and 
thinner solid lines correspond to the 0.5-2 keV band redshift
distributions based on \citet{ueda03} model and \citet{SXLF3}
soft X-ray luminosity function respectively.
\label{fig:zdis}}
\end{inlinefigure}

 An essential ingredient of the de-projection process is the redshift 
distribution of the sources. At this stage, we do not yet have 
individual redshifts of a comprehensive set of the {\sl XMM-Newton} 
complete sample. Thus we use expected distributions from the X-ray luminosity 
functions and AGN population synthesis models. 
We use the model by \citet{ueda03} (Luminosity-dependent density
evolution or LDDE) for all bands. 
In calculating the redshift distribution, we have used the
sensitivity map in units of CR and the actual {\sl XMM-Newton}
response function in each band. We also use \citet{SXLF3} type 1 AGN 
soft X-ray luminosity function (SXLF) for the 0.5-2 keV for comparison. 
The redshift distributions of the  X-ray sources predicted 
by these models are plotted in Fig. \ref{fig:zdis}. 
{Both \citet{ueda03} and \citet{SXLF3} used samples with a very 
high identification completeness with redshifts measurements ($>$ 90\%), 
at least down to the flux limits sampled by the XMM-COSMOS survey. 
Thus the effect on the expected redshift distribution due to the 
identification incompleteness is negligible.}
                                           . 
In calculating the three dimensional correlation functions, we use 
the fits with and without integral constraints. 
The angular correlation amplitude $A$ has been multiplied by a 
correction factor due to PSF merging as discussed in Sect. \ref{sec:merge}. 
Also we use the fits with $\gamma=1.8$.
 The calculated $r_{\rm c,0}$ and the comoving correlation
length at the effective median redshift $r_{\rm c}(\bar{z}_{\rm eff})$ 
are listed in Table~\ref{tab:r0} for selected results.  The errors on 
 $r_{\rm c,0}$ and $r_{\rm c}(\bar{z}_{\rm eff})$ have been calculated for 
fixed $\gamma$ and $\epsilon$. 

\subsection{Bias and Comparison with Other Works}

 In order to estimate the bias parameter of the X-ray sources with respect 
to the underlying mass distribution, we calculate the 
rms fluctuation of the distribution of the X-ray sources in the sphere 
with a comoving radius of $r_{\rm max}=8h^{-1}$ for the power-law model 
\citep[e.g. \S 59 of][]{peebles80},
\begin{eqnarray}
\sigma_{\rm 8,AGN}^2 &=\int \int \xi (|{\bf r_1-r_2}|) dV_1 dV_2/V^2\nonumber\\
     & = (r_{\rm max}/r_{\rm c})^{-\gamma}\; J_2\\
 J_2 & = 72/[(3-\gamma)(4-\gamma)(6-\gamma)2^\gamma].  
\end{eqnarray} 

\begin{figure*}
%\centerline{\includegraphics[width=\textwidth]{sig8bias_col.eps}}
\centerline{\includegraphics[width=\textwidth]{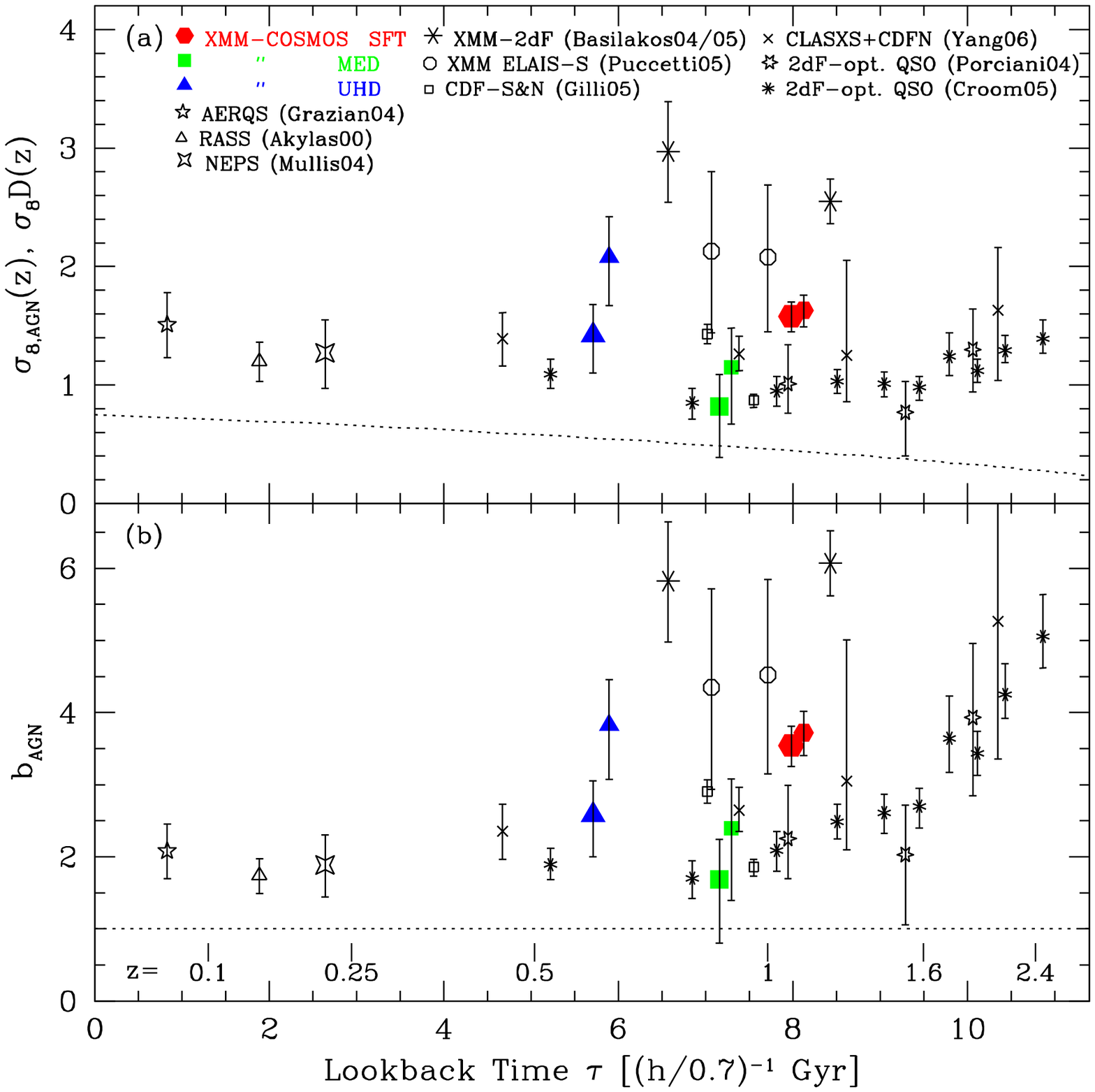}}
\caption
 {(a) The $\sigma_{\rm 8,AGN}$ of the X-ray sources/AGNs inferred by
 the power-law fits to the correlation functions from this work and literature 
 are plotted against the look back time corresponding to the effective median
redshift 
 of the samples. All error bars indicate 1$\sigma$ errors. The results 
 from this work for the fits without integral constraints are plotted with large
solid 
 symbols as labeled (with colors in the electronic version). Those with integral
 constraints are also plotted with smaller symbols and at positions slightly
 shifted rightward for visibility. The $\sigma_{\rm 8,AGN}$ values calculated
 from various results in literature (see text) 
 are also shown as labeled (abbreviated as the first author followed by 
 the last two digits of the publication year). 
 The dotted line shows $\sigma_8\,D(z)$ for the mass in the linear theory 
 normalized to 0.9 at z=0.
 (b) The bias parameter $b_{\rm AGN}=\sigma_{\rm 8,AGN}(z)/[\sigma_8\,D(z)]$ 
  are plotted as a function of the effective redshift. The meaning of the 
  symbols are the same as
  panel (a). We also show redshift ticks at the bottom part of the figure.  
\label{fig:sigbias}}
\end{figure*}

\chg{As discussed above, we have used the results from Fit ID's
S5, S6, M5, M6, U5,and U6, where the fits are made to the angle
range corresponding to $\approx$ 1-16 $h^{-1}$ Mpc at $\bar{z}_{\rm eff}$.} 
The corresponding quantity of the underlying mass distribution at $z=0$,
$\sigma_8$ is one of the commonly used parameters in cosmology 
\citep{spergel03}. In order to compare our results with other similar 
works on a common ground, we calculated $\sigma_{\rm 8,AGN}$ from power-law
representations from literature and plotted them versus the effective redshift
of each sample\footnote{Some authors give the median redshift of
the number distribution of the X-ray sources, while we and some others give
median redshift of the contribution to the clustering signal. We denote the former by 
$\bar{z}$ and the latter by $\bar{z}_{\rm eff}$. We do not make a distinction
between these in Fig. \ref{fig:sigbias}.}. 

For this comparison, we have used the best fit correlation lengths and
slopes ($r_{\rm c},\gamma$) from literature to estimate  $\sigma_{\rm 8,AGN}$
values and their 1$\sigma$ errors. Since each reference has a different
method of presenting results, we take the following strategy 
in calculating  $\sigma_{\rm 8,AGN}$ and its 1$\sigma$ error.
\begin{enumerate}
\item[(a)] If the referenced article gives confidence contours in the 
  ($r_{\rm c},\gamma$) space, we calculate $\sigma_{\rm 8,AGN}$ values for the 
  nominal case as well as at each point in the $L=L_{\rm min}+1$ contour, where 
  $L$ (with the best-fit value $L_{\rm min}$) is either $\chi^2$ or a 
  statistical estimator that varies as $\chi^2$, e.g. Cash C-estimator 
  for \citet{yang06}. The error range on  
   $\sigma_{\rm 8,AGN}$ is determined by the minimum and maximum values  
   calculated from the points along the contour.
\item[(b)] If no confidence contour in the ($r_{\rm c},\gamma$) space is given
   and there is a fit result with a fixed $\gamma$, we use this fit to calculate 
   $\sigma_{\rm 8,AGN}$. The 1$\sigma$ error in $r_{\rm c}$ is propagated from
   that of $r_{\rm c}$.
\item[(c)] If the article gives only best-fit ($r_{\rm c},\gamma$) values, and
   1$\sigma$ errors on both parameters, the error of $\sigma_{\rm 8,AGN}$ has been
   propagated from those of $r_{\rm c}$ and $\gamma$, neglecting possible 
   correlation of errors between  these two parameters. In this case, we may well 
   have over/under estimated the errors on $\sigma_{\rm 8,AGN}$. 
\end{enumerate}
 
 If the ($r_{\rm c},\gamma$) values are given in multiple evolution models,
we use the one where the correlation length is fixed in the comoving coordinates 
(i.e. $\epsilon=-3-\gamma$). At about the effective median redshift 
of the sample, however, the correlation lengths calculated assuming different values
of $\epsilon$ do not differ significantly. In the case of our present work, we see
this by comparing the $r_{\rm c} (\bar{z}_{\rm eff})$ values for $\epsilon=-1.2$ and
$\epsilon=-3$ cases in Table \ref{tab:r0}.  Also the value of $\sigma_{\rm 8,AGN}$ is
insensitive to the assumed value of $\gamma$. The change of $\sigma_{\rm 8,AGN}$ is 
less than 0.1 between the assumed $\gamma$ of 1.8 and 1.5 for our results in all 
the three bands. The results from literature we use for this comparison and the 
details of the conversion to $\sigma_{\rm 8,AGN}$  are described below, roughly
in order of redshift. 

 \citet{grazian04} calculated the correlation function
of 392 optically-selected QSOs from the Asiago-ESO/RASS QSO Survey (AERQS)
with  $\bar{z}_{\rm eff}=0.062$. They found the nominal values of
$r_{\rm c}=8.6\pm 2.0\,h^{-1}$ with $\gamma=1.56$.  
Also in the low-redshift end, \citet{akylas00} calculated the correlation
function of X-ray selected AGNs from the {\it ROSAT} All-Sky Survey (RASS) with 
a median redshift of $\bar{z}=0.15$. 
Their correlation length for  $\gamma=1.8$ of  $r_{\rm c}=6.5\pm 1.0\,h^{-1}$ 
Mpc for the Einstein de-Sitter Universe is increased by 5\% 
to convert it to our adopted cosmology.   
\citet{mullis04} found  $r_{\rm c}=7.4^{+1.8}_{-1.9}\,h^{-1}$ Mpc for $\gamma=1.8$ 
in their {\it ROSAT} North Ecliptic Pole Survey (NEPS) AGNs with median redshift 
for the contribution
to the clustering signal of $\bar{z}_{\rm eff}=0.22$.   
\citet{basilakos04,basilakos05} in their 2 deg$^2$ {\sl XMM-Newton} survey, 
with shallower flux limits than
XMM-COSMOS, found $r_{\rm c}=16.4\pm 1.3\,h^{-1}$ at $\bar{z}=1.2$
and $r_{\rm c}=19\pm 3\,h^{-1}$ at $\bar{z}=0.75$ for the 0.5-2 and 2-8 keV
respectively. {A recent work by \citet{puccetti06} on the central $\sim 0.6$ 
deg$^2$ region of the ELAIS-S1 field , covered by four mosaiced
{\sl XMM-Newton} exposures with $\sim 50-60$ ks each, also measured 
angular ACFs of X-ray point sources. For fixed $\gamma=1.8$, they
found  $r_{\rm c}=12.8\pm 4.2\,h^{-1}$ at $\bar{z}=1.0$
and $r_{\rm c}=17.9\pm 4.8\,h^{-1}$ at $\bar{z}=0.85$ for the 0.5-2 and 2-10
keV bands respectively.}

The correlation functions on the deepest X-ray surveys on the 
{\sl Chandra} Deep Fields- South (CDF-S; $\bar{z}=0.84$) and North 
(CDF-N; $\bar{z}=0.96$) by \citet{gilli05} gave, 
for fixed $\gamma=1.4$, $r_{\rm c}=10.4\pm 0.8\,h^{-1}$ and 
$5.1^{+0.4}_{-0.5}\,h^{-1}$ Mpc respectively. We use the results from their AGN 
samples. For all of the above samples,
the errors on the $\sigma_{\rm 8,AGN}$ have been calculated using method (b).  

An extensive redshift-space correlation function was made by \citet{yang06},
who made use of the data from a combination of the CLASXS and CDF-N surveys, 
with a significant portion of the X-ray sources having measured spectroscopic 
redshifts.  We have used their ($s_0,\gamma$)
confidence contours, where $s_0$ is the redshift-space comoving 
correlation length, in the four redshift bins with median redshifts of
$\bar{z}=$ 0.45, 0.92, 1.26, and 2.07 to estimate  $\sigma_{\rm 8,AGN}$ using method (a).
In this conversion, we have corrected for the redshift distortion by dividing the
redshift-space 
$\sigma_{\rm 8,AGN}$ value by $\sqrt{1.3}$ \citep{marinoni05,yang06}.  Extensive
clustering studies of QSOs from the 2dF QSO redshift survey (2QZ) have been made
using both the projected-distance correlation function approach \citep{porciani04} and
the redshift-space three-dimensional correlation function approach \citep{croom05}. 
We have converted the nominal $r_{\rm 0}$--$\gamma$ values and confidence 
contours in three redshift bins at $\bar{z}_{\rm eff}$=1.06, 1.51 and 1.89 by 
\citet{porciani04} to $\sigma_{\rm 8,AGN}$ using method (a). In converting the 
\citet{croom05}'s ($s_0,\gamma$) results in 10 redshift bins ranging from
$z=0.5$ to 2.5, we have used method c) and the redshift distortion correction
has been made in the same way as we have done to the \citet{yang06} results. 

 Figure \ref{fig:sigbias}(a) shows the $\sigma_{\rm 8,AGN}$ values 
as a function of the look back time $\tau (z)$ for our default cosmology
from our analysis results both without and with integral constraints. 
We also overplot  $\sigma_{\rm 8,AGN}$ values calculated from the results found
in literature as detailed above.   In order to compare them 
with those of the underlying mass distribution, we have also plotted the 
$\sigma_{\rm 8}\,D(z)$ values from the linear theory
\citep[e.g.][]{carroll92,hamilton01}, 
normalized to 0.75 at z=0 \citep{spergel03}
\footnote{We use the latest value of $\sigma_8$ as of writing this paper obtained from \\  
\url{http://lambda.gsfc.nasa.gov/product/map/current/parameters.cfm } 
for the $\Lambda$-CDM model derived from all datasets.}. 
This curve has been shown to accurately represent the distribution of dark 
matter particles in the $\Lambda$CDM Hubble Volume 
Simulation \citep{marinoni05}. Figure \ref{fig:sigbias}(b) shows the inferred
bias parameters $b_{\rm AGN}=\sigma_{\rm 8,AGN}(z)/[\sigma_8\,D(z)]$. The values
of $\sigma_{\rm 8,AGN}$ and $b_{\rm AGN}$ from this work are shown in 
Table \ref{tab:sigbias}. 

\section{Discussion and Prospects} 
\label{sec:disc}

 In this work, we used all the point sources above the scaled sensitivity
threshold without further classification of the sources. We analyzed 
our results assuming that all the X-ray sources are AGNs. This, in practice,
is a good approximation. Our preliminary identifications of the sources 
indicate that out of the 1037, 545, and 151 sources selected for the ACF 
analysis for the SFT, MED and UHD bands, 20, 5, and 1 are apparent Galactic 
stars respectively. Removing 
these sources from the analysis changed the results very little. Also our
results 
are not likely to be heavily affected by the contamination of clusters/groups,
since these
sources are extended by $\gtrsim 20\arcsec$ \citep[e.g.][]{finoguenov06} and
are likely to be classified as extended by the source detection procedure, 
hence removed from our sample. 

 As seen in Fig. \ref{fig:sigbias}, \chg{with an exception of
the MED band,} our analysis without integral constraints 
gives somewhat larger $\sigma_{\rm 8,AGN}$ values than those obtained from
results 
using 2dF optically selected QSOs by \citet{porciani04} and \citet{croom05}, 
but in general agreement with the values from {\sl Chandra} CLASXS+CDF-N by 
\citet{yang06}, CDF-S by \citet{gilli05}, {and {\sl XMM-Newton} results
from \citet{puccetti06}}. Most likely due to the cosmic
variance over a small FOV, \citet{gilli05}'s result on CDF-N gave 
a significantly smaller correlation amplitude than their own CDF-S values as
well as our results. The angular ACFs from a shallower 
{\sl XMM-Newton} survey by \citet{basilakos04,basilakos05} gave 
significantly larger $\sigma_{\rm 8,AGN}$ values than other works in both
0.5-2 keV and 2-8 keV bands. The reason for their distinctively large 
value is unclear.

 \chg{One of the interesting questions in investigating clustering
properties of X-ray selected AGNs is to investigate whether there
is any difference in the environments of obscured and unobscured 
AGNs.} Applying the population synthesis model of \citet{ueda03} to our 
sensitivity maps, only $\sim 17$\% of the sources detected in the SFT 
band at $z\sim 1$ are obscured AGNs with $N_{\rm H}>10^{22}$ cm$^{-2}$. 
The fraction increases to $\sim 40$\% in the MED and UHD bands. 
\chg{ A comparison of bias parameters between SFT band and MED band, 
which have similar $\bar{z}_{\rm eff}$ values, seems to
show a lager bias parameter  for the SFT sample. However, with
the combination of statistical uncertainties and uncertainties in 
modeling the integral constraint in the MED band, we can only
conclude that the bias of the obscured AGNs is not 
stronger than that of unobscured AGNs. 
In other works, \citet{gilli05,yang06} as well as \citet{basilakos04,basilakos05}
did not find any statistical difference between the clustering properties
of these two.} Further studies involving 
{the second-year XMM-COSMOS data, which in effect doubles the 
{\sl XMM-Newton} exposure over the COSMOS field,} and redshift 
information of individual objects  will probe into this problem further.
\chg{Also with the accepted C-COSMOS program totaling 1.8 Ms of {\sl Chandra}
exposure, we will be able to  probe the correlation functions to a much smaller 
scale, enabling us to investigate the immediate neighbor environments of these AGNs.}

 Our measured bias parameters based on the rms fluctuations in the 8$h^{-1}$ Mpc 
radius sphere are in the range $b_{\rm AGN}=1.5-4$. The clustering properties of dark 
matter halos (DMH) depend on their mass \citep{mo96,sheth01} and we can estimate 
the typical mass of the DMHs in which the population of AGNs represented by our
sample  reside, under the assumption that the typical mass halo is the main cause of
the AGN biasing. Following the approach of \citet{yang06} and \citet{croom05} who utilized 
the model by \citet{sheth01}, we roughly estimate that 
the typical mass of DMH is $\sim 10^{13}-10^{14}$ $M_\sun$ for \chg{our SFT and UHD samples} 
(see Table \ref{tab:sigbias}). These are an order of
magnitude larger than those estimated by \citet{porciani04} and \citet{croom05}, probably 
reflecting the large bias parameters from our results. 

One of the largest uncertainties in our analysis lies in the 
treatment of the integral constraint, because its effect is not
negligible in our case compared with the ACF amplitudes in the range
of our interest. Fig. \ref{fig:sigbias} shows that our results based on the
fits with integral constraint, under an assumption that the 
fitted power-law behavior of the underlying $w(\theta)$ extends to the scale of 
the entire FOV, give a somewhat larger correlation amplitudes. 
This assumption may not be true. Also the apparent negative 
$w_{\rm est}(\theta)$ values at $\theta >30^{\arcmin}$ in Fig. \ref{fig:acf3}
may well be caused by remaining systematic errors. Thus the interpretation of 
the angular correlation functions, where the signals are diluted by the projection 
along the redshift space, has a major limitation in correctly taking the integral constraint
into account.

 The situation will improve when redshift information
on individual X-ray sources becomes available for a major and 
comprehensive set of the X-ray sources. This will enable us
to calculate three-dimensional correlation functions 
or projected-distance correlation functions in a number 
of redshift bins. With the line-of-sight dilution effect suppressed, 
we will be able to obtain a larger amplitude in the correlation signal, making
the analysis much less subject to the uncertainties in the 
integral constraint. With optical followup programs underway on the COSMOS field
through Magellan and zCOSMOS projects \citep{trump06,lilly06}, we are obtaining 
spectroscopic redshifts from a major fraction of the X-ray sources.
At the time of writing this paper, we have been able to define a sample of 
378 XMM-COSMOS detected AGNs with measured spectroscopic redshifts 
($\sim 30$\% of the X-ray point sources), with a median 
redshift of $z\sim 1$. Our preliminary analysis of these sources based on the  
projected-distance correlation function $w(r_{\rm p})$ gives a comoving
correlation length of $r_{\rm c}\approx 8h^{-1}$ Mpc and $\gamma=1.6$, 
which is fully consistent with our results without the integral constraints. 
The results of a full analysis utilizing the redshift information will be 
presented in a future paper (Gilli et al. in preparation).
  
 Extensive multi-wavelength coverage and the availability of a galaxy catalog
also enables us to investigate the cross-correlation function between X-ray 
selected AGNs and galaxies. 
By cross-correlating the X-ray selected AGNs with three orders of magnitude larger
number of galaxies, we will be able to investigate the environments of the AGN 
activity in various redshifts with much better statistics.

\section{Conclusions}
\label{sec:conc}

We have presented the first results on the angular correlation functions
(ACFs) of the X-ray selected AGNs from the XMM-COSMOS survey and reached the 
following conclusions.
\begin{enumerate}
\item \chg{A significant positive angular clustering signals has been detected 
   in the 0.5-2 (SFT) bands in the angle range of $0\farcm 5$-20$\arcmin$, while 
   in the  2-4.5 (MED) and 4.5-10 keV  (UHD) bands, the positive signals are
   2 and 3$\sigma$ respectively.} The robustness of the estimated correlation 
   functions has been verified using different methods of generating random samples.
\item Power-law fits to the angular correlation function have been made, taking
   into account the correlation of errors. Correctly taking the integral constraint
   into account is a major limitation on interpreting the angular correlation
   function. \chg{For fits with fixed $\gamma-1=0.8$
   and without (with) the integral constraint term, we found 
   correlation lengths of \chg{$\theta_{\rm c}=1\farcs 9\pm 0\farcs 3$,  
    $0\farcs 8^{+0\farcs 5}_{-0\farcs 4}$ 
   and $6\arcsec \pm 2\arcsec$ ($3\farcs 1\pm 0\farcs 5$, $2\farcs 2\pm 1\farcs 0$, and  
   $14\arcsec \pm 5\arcsec$)} for the SFT, MED, and UHD bands respectively.}  
\item Due to {\sl XMM-Newton} PSF, most of the source pairs closer than $\sim 20\arcsec$ 
   are classified as single extended sources, and therefore excluded from the
   sample. This 
   causes a bias in angular correlation function measurements. We have estimated this 
   effect (the PSF merging bias) by simulations and found that the estimated ACF 
   underestimates the amplitude of 
   the true underlying ACF by $\sim 15$\% and $\sim 8$\% 
   for the SFT and MED bands respectively  
\item Using Limber's equation and the expected redshift distributions of the
   sources, we have found comoving correlation lengths of 
   $r_{\rm c}\approx 9.8\pm 0.7$, 
   $5.8^{+1.4}_{-1.7}$, and $12\pm 2$ $h^{-1}$ Mpc for $\gamma=1.8$ at
   the effective redshifts of $\bar{z}_{\rm eff}\approx$ 1.1, 0.9, and 0.55 for
   the  SFT, MED, and UHD bands respectively for the fits without integral
   constraints, while 20\%-90\%  larger correlation lengths have been obtained for the fits
   with integral constraints.
\item Using the fits in the angles corresponding to a projected distance
    range of 1-16$h^{-1}$ Mpc at the effective median redshift of the sample, 
    we have calculated the rms fluctuations of the X-ray source distributions. 
    Comparing them with that of the mass distribution from the linear theory, 
    we find that the bias parameters of the X-ray 
    sources are in the range $b_{\rm AGN}=1.5-4$ at $0.5<z<1.2$. 
%\item Currently no indication  is found for different clustering properties between
%   obscured and unobscured AGNs in our data. 
\item If the bias mainly reflects the typical mass of dark matter halos in which
    these X-ray AGNs reside, their typical masses are $10^{13}-10^{14}$ $M_\sun$. 
\item Further investigations utilizing redshifts of individual X-ray sources
  and/or involving cross-correlation function with galaxies taking advantage of the
  wealth of multiwavelength data are being conducted. \chg{The approved {\sl Chandra} 
  observations (C-COSMOS) on this field will enable us to probe into the clustering in much
   smaller scales and therefore into immediately neighboring environments of
   AGNs.}
\end{enumerate}

\acknowledgments
This work is based on observations obtained with XMM-Newton, an
ESA science mission, with instruments and contributions directly funded 
by ESA member states and the US (NASA). Partial support of this work is provided by
NASA LTSA Grant NAG5-10875 (T.M.) and the US {\it XMM-Newton} 
Guest Observer Support NNG04GG40G/NNG06GD77G (T.M.,R.E.G.).  
In Germany, the XMM--Newton project is supported
by the Bundesministerium f\"ur Bildung und Forschung/Deutsches Zentrum f\"ur Luft und
Raumfahrt, the Max--Planck Society, and the Heidenhain--Stiftung. Part of this
work was supported by the Deutsches Zentrum f\"ur Luft-- und Raumfahrt,
DLR project numbers 50 OR 0207 and 50 OR 0405. In Italy, the XMM-COSMOS 
project is supported by INAF and MIUR under grants PRIN/270/2003 and 
Cofin-03-02-23. The COSMOS Science meeting in May 2005 was supported in part by
the NSF through grant OISE-0456439. We gratefully acknowledge the contributions 
of the entire COSMOS collaboration consisting of more than 70 scientists. In particular,
we thank Manfred Kitbichler for allowing us to use his COSMOS-Mock catalogs and 
Chris Carilli for reviewing the manuscript. More information on the COSMOS survey is 
available at {\bf \url{http://www.astro.caltech.edu/~cosmos}}. It is a 
pleasure the acknowledge the excellent services provided by the NASA IPAC/IRSA
staff (Anastasia Laity, Anastasia Alexov, Bruce Berriman and John Good)
in providing online archive and server capabilities for the COSMOS datasets.

{\it Facilities:} \facility{XMM ()}

\begin{deluxetable}{ccccc}
\footnotesize
\tablecaption{X-ray sources and Sensitivity limits\label{tab:src}}
\tablewidth{0pt}
\tablehead{
\colhead{Band} & 
\colhead{Number} & 
\colhead{$CR_{\rm lim,min}$--$CR_{\rm lim,max}$} & 
\colhead{$S_{\rm x,lim}$ range} &
\colhead{Area}   \\
\colhead{(keV)}&                        
& 
\colhead{(cts\,s$^{-1}$)} &
\colhead{(erg s$^{-1}$ cm$^{-2}$)\tablenotemark{a}} &
\colhead{(deg$^{-2}$)} 
}
\startdata
SFT  & 1037 & $7.0\;10^{-4}$--$2.2\; 10^{-2}$ & $6.7\;10^{-16}$--$2.1\;10^{-14}$
& 1.43 \\
MED  &  545 & $7.0\;10^{-4}$--$2.5\; 10^{-2}$ & $4.6\;10^{-15}$--$1.6\;10^{-13}$
& 1.56 \\
UHD  &  151 & $9.0\; 10^{-4}$--$2.2\; 10^{-2}$ &
$8.7\;10^{-15}$--$1.8\;10^{-13}$ & 1.25\\ 
\enddata
\tablenotetext{a}{Flux range corresponding to the $CR$ limits.
The conversions have been made, following C06, 
to fluxes in 0.5-2.0, 2.0-10, 5.0-10 keV 
assuming power-law spectra of photon indices 
$\Gamma=2.0$, 1.7, and 1.7 for the SFT, MED, and UHD bands respectively.} 
\end{deluxetable}

\begin{deluxetable}{cccccccc}
\footnotesize
\tablecaption{Results of the Power-law fits\label{tab:fit}}
\tablewidth{0pt}
\tablehead{
\colhead{Fit ID} &
\colhead{Band} &
\colhead{$A$\tablenotemark{a}} & 
\colhead{$\gamma-1$} & 
\colhead{$\theta_{\rm c}$\tablenotemark{a}} &
\colhead{$C$} &
\colhead{$\theta_{\rm min}$} & 
\colhead{$\theta_{\rm max}$} 
\\
&
\colhead{(keV)}&                        
\colhead{(arcmin$^{1/(\gamma-1)}$)} &
&
\colhead{(\arcsec)} &
&
\colhead{(\arcmin)} &
\colhead{(\arcmin)}
}
\startdata
%                A          gamma-1    thc          C   thmin thmax
S1 & SFT & $0.063\pm .008$ & $0.8$ & $1.9\pm 0.3$ & 0 & 0.5 & 24\\
S2 & SFT & $0.093\pm .012$ & $0.8$ & $3.1\pm 0.5$ & $8\;10^{-3}$ & 0.5 & 24\\
S3 & SFT & $0.034\pm .004$ & $0.5$ & $0.07\pm .02$ & 0 & 0.5 & 24\\
S4 & SFT & $0.078\pm .010$ & $0.5$ & $0.37\pm .09$ & 1.5$\; 10^{-2}$ & 0.5 & 24\\
S5 & SFT & $0.059\pm .009$ & $0.8$ & $1.7\pm 0.3$ & 0 & 1.6 & 24 \\       
S6 & SFT & $0.089\pm .015$ & $0.8$ & $2.9\pm 0.6$ & $7\;10^{-3}$  & 1.6 & 24 \\
\\
M1 & MED & $0.032\pm .015$ & $0.8$ & $0.8^{+0.5}_{-0.4}$ & 0 & 0.7 & 24\\
M2 & MED & $0.071\pm .027$ & $0.8$ & $2.2\pm 1.0$ & 5$\; 10^{-3}$ & 0.7 & 24\\
M3 & MED & $0.013\pm .008$ & $0.5$ & $.010^{+.016}_{-.009}$ & 0 & 0.7 & 24\\
M4 & MED & $0.048\pm .020$ & $0.5$ & $.14^{+.14}_{-.09}$ & 9$\; 10^{-3}$ & 0.7 & 24\\
M5 & MED & $0.021\pm .016$ & $0.8$ & $.47^{+.49}_{-.40}$ & 0   & 1.6 & 24\\       
M6 & MED & $0.044\pm .029$ & $0.8$ & $1.2^{+1.1}_{-0.9}$ & 3$\;10^{-3}$ & 1.6 & 24\\
\\
U1 & UHD & $0.15\pm .05$   & $0.8$ & $5.6\pm 2.3$ & 0 & 0.5 & 24\\
U2 & UHD & $0.32\pm .09$   & $0.8$ & $14\pm$ 5 & 3$\; 10^{-2}$ & 0.7 & 24\\   
U3 & UHD & $0.075\pm .024$ & $0.5$ & $0.34^{+0.25}_{-0.18}$ & 0 & 0.7 & 24\\
U4 & UHD & $0.23\pm .07$   & $0.5$ & $3.2^{+2.2}_{-1.6}$ & 5$\; 10^{-2}$ & 0.7 & 24\\ 
U5 & UHD & $0.080\pm .032$ & $0.8$ & $2.5^{+1.3}_{-1.2}$ &    0  & 2.4 & 35 \\       
U6 & UHD & $0.17\pm .06$   & $0.8$ & $6.5^{+3.0}_{-2.7}$ & 3$\; 10^{-2}$ & 2.4 & 35 \\
\enddata
\tablenotetext{a}{One $\sigma$ errors are shown. The effects of PSF merging
(Sect. \ref{sec:merge}) have not been taken into account.}
\end{deluxetable}

\begin{deluxetable}{ccccccc}
\footnotesize
\tablecaption{Three Dimensional Correlation Lengths\label{tab:r0}}
\tablewidth{0pt}
\tablehead{
\colhead{Fit ID} &
\colhead{$\gamma$} &
\colhead{$\epsilon$}&  
\colhead{$\bar{z} _{\rm eff}$}& 
\colhead{$r_{\rm c,0}$\tablenotemark{a}} &
\colhead{$r_{\rm c} (\bar{z}_{\rm eff})$\tablenotemark{a}} &
\colhead{Model\tablenotemark{b}} 
\\
&
& 
&
&
\multicolumn{2}{c}{[$h^{-1}$ Mpc]}& 
}
\startdata
S1 & 1.8 & -1.2 & 1.07 & 9.8$\pm 0.7$  & 9.8$\pm 0.7$ & U03 \\  
S1 & 1.8 & -1.2 & 1.11 & 9.4$\pm 0.7$  & 9.4$\pm 0.7$ & H05 \\
S1 & 1.8 & -3.0 & 1.42 & 4.3$\pm 0.3$  & 10.4$\pm 0.7$ & U03 \\
S2 & 1.8 & -1.2 & 1.07 & 12.1$\pm 0.9$  & 12.1$\pm 0.9$ & U03 \\
S5 & 1.8 & -1.2 & 1.07 & 9.4$\pm 0.8$  & 9.4$\pm 0.8$ & U03 \\
S6 & 1.8 & -1.2 & 1.07 & 11.8$\pm$1.1  & 11.8$\pm$1.1 & U03 \\
M1 & 1.8 & -1.2 & 0.87 & 5.8$^{+1.4}_{-1.7}$ & 5.8$^{+1.4}_{-1.7}$ & U03 \\
M1 & 1.8 & -3.0 & 1.13 & 2.9$^{+0.7}_{-0.8}$ & 6.1$^{+1.5}_{-1.8}$ & U03 \\
M2 & 1.8 & -1.2 & 0.87 & 9.0$^{+1.8}_{-2.1}$ & 9.0$^{+1.8}_{-2.1}$ & U03 \\
M5 & 1.8 & -1.2 & 0.87 & 4.6$^{+1.7}_{-2.5}$ & 4.6$^{+1.7}_{-2.5}$ & U03 \\
M6 & 1.8 & -1.2 & 0.87 & 6.9$^{+2.2}_{-3.1}$ & 6.9$^{+2.2}_{-3.1}$ & U03 \\
U1 & 1.8 & -1.2 & 0.60 & 11.9$^{+2.1}_{-2.4}$ & 11.9$^{+2.1}_{-2.4}$ & U03\\
U1 & 1.8 & -3.0 & 0.88 & 6.6$^{+1.1}_{-1.2}$ & 12.5$^{+2.2}_{-2.5}$  & U03\\
U2 & 1.8 & -1.2 & 0.60 & 19$\pm 3$ & 19$\pm 3$ & U03\\
U5 & 1.8 & -1.2 & 1.60 & 8.4$^{+1.7}_{-2.0}$ & 8.4$^{+1.7}_{-2.0}$   & U03 \\
U6 & 1.8 & -1.2 & 1.60 & 12.7$^{+2.3}_{-2.7}$ & 12.7$^{+2.3}_{-2.7}$ & U03 \\
\enddata
\tablenotetext{a}{
Errors are 1 $\sigma$.}
\tablenotetext{b}{Model designations-- U03:\citet{ueda03}, H05:\citet{SXLF3}}
\end{deluxetable}

\begin{deluxetable}{ccccc}
\footnotesize
\tablecaption{Estimated $\sigma_{\rm 8,AGN}$ and Bias\label{tab:sigbias}}
\tablewidth{0pt}
\tablehead{
\colhead{Fit ID} &
\colhead{$\bar{z}_{\rm eff}$}& 
\colhead{$\sigma_{\rm 8,AGN}$} &
\colhead{$b_{\rm AGN}$} &
\colhead{$\log M_{\rm halo}$\tablenotemark{a}}\\
&
&
&
&
$h^{-1}\,M_{\sun}$ 
}
\startdata
\multicolumn{5}{l}{{\em Without integral constraint\ldots}}\\
S5 & 1.07 & 1.58$^{+.12}_{-.13}$ & 3.5$\pm$0.3          & 13.6$\pm 0.1$\\
M5 & 0.87 & 0.82$^{+.27}_{-.42}$ & 1.7$^{+0.6}_{-0.9}$  & $<$13.3 \\
U5 & 0.60 & 1.42$^{+.26}_{-.32}$ & 2.6$^{+0.5}_{-0.6}$  & 13.5$\pm 0.2$\\
\tableline
\multicolumn{5}{l}{{\em With integral constraint\ldots}}\\
S6 & 1.07 & 1.63$^{+.13}_{-.14}$ & 3.7$\pm$0.3 & 13.6$\pm$0.1\\
M6 & 0.87 & 1.19$^{+.34}_{-.50}$ & 2.5$^{+0.7}_{-1.0}$  & 13.3$^{+0.3}_{-0.7}$\\
U6 & 0.60 & 2.08$^{+.34}_{-.41}$ & 3.8$^{+0.6}_{-0.8}$  & 13.9$\pm 0.2$\\
\enddata
\tablenotetext{a}{The error on $M_{\rm halo}$ reflects the statistical error on
$b_{\rm AGN}$ only.}
\end{deluxetable}

\end{document}